\documentclass[12pt]{article}
\usepackage{amssymb, amsmath, amsfonts}

\newtheorem{thm}{Theorem}[section]

\newtheorem{cor}[thm]{Corollary}
\newtheorem{lem}[thm]{Lemma}

\font\myfont=msbm10 scaled \magstep1

\def\ZZ{\hbox{\myfont\char'132}}
\def\NN{\hbox{\myfont\char'116}}

\begin{document}

\begin{center}
{\Large Complete list of Darboux Integrable Chains of the form
$t_{1x}=t_x+d(t,t_1)$}

\vskip 0.2cm

{Ismagil Habibullin}\footnote{e-mail: habibullinismagil@gmail.com}\\

{Ufa Institute of Mathematics, Russian Academy of Science,\\
Chernyshevskii Str., 112, Ufa, 450077, Russia}\\
\bigskip

{Natalya Zheltukhina}

{Asl{\i} Pekcan}

{Department of Mathematics, Faculty of Science,
 \\Bilkent University, 06800, Ankara, Turkey \\}

\end{center}

\begin{abstract}
We study differential-difference equation of the form
$$
\frac{d}{dx}t(n+1,x)=f(t(n,x),t(n+1,x),\frac{d}{dx}t(n,x))
$$
with unknown $t(n,x)$ depending on continuous  and discrete
variables $x$ and  $n$. Equation of such kind is called Darboux
integrable, if there exist two functions $F$  and $I$ of a finite
number of arguments $x$, $\{t(n\pm k,x)\}_{k=-\infty}^\infty$,
$\left\{\frac{d^k}{dx^k}t(n,x)\right\}_{k=1}^\infty$,
 such that $D_xF=0$ and $DI=I$, where $D_x$
is the operator of total differentiation with respect to $x$, and
$D$ is the shift operator: $Dp(n)=p(n+1)$.
 Reformulation of  Darboux
integrability  in terms of finiteness of two
characteristic Lie algebras  gives an effective tool for
classification of integrable equations. The complete list of Darboux
integrable
equations  is given in
the case when the function $f$ is of the special form
$f(u,v,w)=w+g(u,v)$.
\end{abstract}

{\it Keywords:} semi-discrete chain, classification, $x$-integral,
$n$-integral, characteristic Lie algebra, integrability
conditions.

\section{Introduction}

In this paper we  continue investigation of integrable
semi-discrete chains of the form
\begin{equation}\label{dhyp}
\frac{d}{dx}t(n+1, x)=f(t(n, x),t(n+1, x),\frac{d}{dx}t(n, x)),
\end{equation}
started in our previous paper \cite{HZhP}. Here $t=t(n,x)$ and
$t_1=t(n+1,x)$ are unknown. Function $f=f(t,t_1,t_x)$ is assumed to
be locally analytic and $\frac{\partial f}{\partial t_x}$ is not
identically zero. Nowadays discrete phenomena are very popular due
to their applications in physics, geometry, biology,  etc. (see,
\cite{Zabrodin}-\cite{GKP} and references therein).

Below we use subindex to indicate the shift of the discrete
argument: $t_k=t(n+k,x)$, $k\in \ZZ$,   and derivatives with respect
to $x$: $t_{[1]}=t_x=\displaystyle{\frac{d}{d x}}t(n,x),$
$t_{[2]}=t_{xx}=\displaystyle{\frac{d^2}{d x^2}}t(n,x)$,
$t_{[m]}=\frac{d^m}{dx^m}t(n,x)$, $m\in \NN$.  Introduce the set of
dynamical variables containing $\{t_k\}_{k=-\infty}^{\infty};$
$\{t_{[m]}\}_{m=1}^{\infty}$.

We denote through $D$ and $D_x$ the shift operator and the
operator of the total derivative with respect to $x$ correspondingly. For
instance, $Dh(n,x)=h(n+1,x)$ and
$D_xh(n,x)=\frac{d}{d x}h(n,x)$.

Functions $I$ and $F$, both depending on $x$ and a finite number of
dynamical variables, are called  respectively $n$- and $x$-integrals
of (\ref{dhyp}), if  $DI=I$ and $D_xF=0$ (see also
\cite{AdlerStartsev}). One can see that any $n$-integral $I$ does
not depend on variables $t_{m}$, $m\in \ZZ\backslash\{0\}$
 and any $x$-integral $F$
does not depend on variables $t_{[m]}$, $m\in \NN$.

Chain (\ref{dhyp}) is called Darboux integrable
if it admits a nontrivial $n$-integral  and a nontrivial
$x$-integral.

Note that all Darboux integrable chains of the form (\ref{dhyp}) are
reduced to the d'Alembert equation $w_{1x}-w_x=0$ by the following
"differential" substitution $w=F+I$. Indeed,
$D_x(D-1)w=(D-1)D_xF+D_x(D-1)I=0$. This implies that two arbitrary
Darboux integrable chains of the form (\ref{dhyp})
$$
u_{1x}=f(u,u_1,u_x), \quad v_{1x}=\tilde f(v,v_1,v_x)
$$
are connected with one another by the substitution
$$
F(x,u,u_1,u_{-1},...)+I(x,u,u_x,u_{xx},...)=\tilde F(x,v,v_1,v_{-1},...)+\tilde I(x,v,v_x,v_{xx},...),
$$
which is evidently splitted down into two relations
$$
F(x,u,u_1,u_{-1},...)=\tilde F(x,v,v_1,v_{-1},...)-h, \quad
I(x,u,u_x,u_{xx},...)=\tilde I(x,v,v_x,v_{xx},...)+h,
$$
where $h$ is some constant.

The idea of such kind integrability goes back to Laplace's discovery
of cascade method of integration of linear hyperbolic type PDE with
variable coefficients  made in 1773 (see \cite{Laplace}). Roughly speaking the Laplace
theorem claims that a linear hyperbolic PDE admits general solution
in a closed form if and only if its sequence of Laplace invariants
terminates at both ends (see \cite{Darboux}). More than hundred
years later G.~Darboux applied the cascade method to the nonlinear
case. He proved that a nonlinear hyperbolic equation is integrable
(Darboux integrable) if and only if the Laplace sequence of the
linearized equation terminates at both ends. This result has been
rediscovered very recently by I.~M.~Anderson, N.~Kamran
\cite{AndersonKamran}, and V.~V.~Sokolov, A.~V.~Zhiber
\cite{Zhiber}.

An alternative approach was suggested by A.~B.~Shabat and R.~I.~Yamilov in 1981 (see \cite{ShabatYamilov}). They assigned two Lie
algebras, called characteristic Lie algebras to each hyperbolic
equation and proved that the equation is Darboux integrable if and
only if both characteristic Lie algebras are of finite dimension.

The purpose of the present article is to study characteristic Lie
algebras of the chain (\ref{dhyp}) introduced in our papers
\cite{Habibullin}-\cite{TJM}  and convince the reader that in
the discrete case these algebras provide a very effective
classification tool.

We denote through $L_x$ and $L_n$ characteristic Lie algebras in $x$-
and $n$-directions, respectively. Remind the definition of $L_x$.
Rewrite first the chain (\ref{dhyp}) in the inverse form
$t_x(n-1, x)=g(t(n, x),t(n-1, x),t_x(n, x))$. It can be done (at least locally) due to the
requirement $\frac{\partial f}{\partial t_x}(t,t_1,t_x)\neq 0$. An
$x$-integral $F=F(x,t,t_{\pm 1},t_{\pm 2},...)$ solves the
equation $D_xF=0$. Applying the chain rule, one gets $KF=0$, where
\begin{equation}\label{gc1} K = \frac{\partial }{\partial x}+t_x\frac{\partial
}{\partial t} +f\frac{\partial }{\partial t_1 }+g\frac{\partial
}{\partial t_{-1}} +f_1\frac{\partial }{\partial
t_2}+g_{-1}\frac{\partial }{\partial t_{-2} }+\ldots\, .
\end{equation}
Since  $F$ does not depend on the variable $t_x$, then $XF=0$,
where $X=\frac{\partial}{\partial t_x}$. Therefore, any vector field
from the Lie algebra generated by $K$ and $X$ annulates $F$. This
algebra is called the characteristic Lie algebra $L_x$ of chain
(\ref{dhyp}) in $x$-direction. The notion of characteristic algebra
is very important. One can prove that chain (\ref{dhyp}) admits a
nontrivial $x$-integral if and only if its Lie algebra $L_x$ is of
finite dimension. The proof of the next classification Theorem from
\cite{HZhP} is based  on the finiteness of the Lie algebra $L_x$.
\begin{thm}\label{thm}
Chain
\begin{equation}\label{main}
t_{1x}=t_x+d(t,t_1)
\end{equation}
admits a nontrivial x-integral if and only if $d(t,t_1)$ is one of
the following kinds;
\begin{enumerate}
\item[(1)] $d(t,t_1)=A(t_1-t)$, \item[(2)]
$d(t,t_1)=c_1(t_1-t)t+c_2(t_1-t)^2+c_3(t_1-t)$, \item[(3)]
$d(t,t_1)=A(t_1-t)e^{\alpha t}$, \item[(4)]
$d(t,t_1)=c_4(e^{\alpha t_1}-e^{\alpha t})+c_5(e^{-\alpha
t_1}-e^{-\alpha t})$,
\end{enumerate}
\noindent where $A=A(t_1-t)$ is an arbitrary function of one
variable and $\alpha\ne 0$, $c_1\ne 0$, $c_2$, $c_3$, $c_4\ne 0$,
$c_5\ne 0$ are arbitrary constants. Moreover, some nontrivial
$x$-integrals in each of the cases are
\begin{enumerate}
\item[i)] $F=x+\int^{t_1-t}\frac{du}{A(u)}$,\quad if \quad $A(u)\neq 0,$\\
$F=t_1-t$,\quad if\quad $A(u)\equiv0$,\\
 \item[ii)]
$F=\frac{1}{(-c_2+c_1)}\ln{|(-c_2+c_1)\frac{t_2-t_1}{t_3-t_2}+c_2|}+
\frac{1}{c_2}\ln{|c_2\frac{t_2-t_1}{t_1-t}-c_2+c_1|}$\, for $\,c_2(c_2-c_1)\neq0,$\\

$F=\ln{\left|\frac{t_2-t_1}{t_3-t_2}\right|}+\frac{t_2-t_1}{t_1-t}$ for $\,c_2=0,$\\

$F=\frac{t_2-t_1}{t_3-t_2}+\ln{\left|\frac{t_2-t_1}{t_1-t}\right|}$ for $\,c_2=c_1,$ \\
\item[iii)] $F=\int^{t_1-t}\frac{\mathrm{e}^{-\alpha u}du}{A(u)}-\int^{t_2-t_1}\frac{du}{A(u)}$,\\
\item[iv)] $F=\frac{(\mathrm{e}^{\alpha t}-\mathrm{e}^{\alpha
t_2})(\mathrm{e}^{\alpha t_1}-\mathrm{e}^{\alpha
t_3})}{(\mathrm{e}^{\alpha t}-\mathrm{e}^{\alpha
t_3})(\mathrm{e}^{\alpha t_1}-\mathrm{e}^{\alpha t_2})}$.
\end{enumerate}
\end{thm}

In what follows we study semi-discrete chains (\ref{main}) admitting
not only nontrivial $x$-integrals but also nontrivial
$n$-integrals. First of all we will give an equivalent algebraic
formulation of the $n$-integral existence problem. Rewrite the
equation $DI=I$ defining $n$-integral in an enlarged form
\begin{equation}\label{n-integral}
I(x,t_1,f,f_{x},...)=I(x,t,t_x,t_{xx},...).
\end{equation}
The left hand side contains the variable $t_1$ while the right
hand side does not. Hence we have $D^{-1}\frac{d}{dt_1}DI=0,$ i.e.
the $n$-integral is in the kernel of the operator
$$
Y_1=D^{-1}Y_0D, $$ where
\begin{equation}\label{Y1}
    Y_1=\frac{\partial}{\partial t}+D^{-1}(Y_0f)\frac{\partial}{\partial t_x}+D^{-1}Y_0(f_x)\frac{\partial}{\partial t_{xx}}+   D^{-1}Y_0(f_{xx})\frac{\partial}{\partial t_{xxx}}+...,
\end{equation}
and \begin{equation} Y_0=\frac{d}{dt_1}\,.
\end{equation}
 It can easily be shown
that for any natural $j$ the equation  $D^{-j}Y_0D^jI=0$ holds.
Direct calculations show that
$$
D^{-j}Y_0D^j=X_{j-1}+Y_j, \qquad j\geq 2,
$$
where
\begin{eqnarray}
&&Y_{j+1}=D^{-1}(Y_jf)\frac{\partial}{\partial t_x}
+D^{-1}Y_j(f_x)\frac{\partial}{\partial t_{xx}}+
D^{-1}Y_j(f_{xx})\frac{\partial}{\partial t_{xxx}}+...,\quad j\geq
1\, , \label{Yj}
\end{eqnarray}
\begin{equation}\label{definitionXj}
X_j=\frac{\partial}{\partial_{t_{-j}}}, \qquad  j\geq
1.\end{equation}
The following theorem defines the characteristic
Lie algebra $L_n$ of the chain (\ref{dhyp}).

\begin{thm}\label{thm1}\cite{HabibullinPekcan}
Equation (\ref{dhyp}) admits a nontrivial $n$-integral if and only
if the following two conditions hold:\\
1)  Linear space spanned by  the operators $\{Y_j \}_1^{\infty}$
is
of finite dimension, denote this dimension by $N$;\\
2)  Lie algebra $L_n$ generated by the operators
${Y_1,Y_2,...,Y_N,X_1,X_2,...,X_N}$ is of finite dimension. We
call $L_n$ the characteristic Lie algebra of (\ref{dhyp}) in the
direction of $n$.
\end{thm}

We use $x$-integral classification Theorem \ref{thm} and $n$-integral existence Theorem \ref{thm1}
to obtain the complete list of
Darboux integrable chains of the form  (\ref{main}).
The statement of this  main result of the present paper is given in the next   Theorem.

\begin{thm}\label{maintheorem}
Chain (\ref{main}) admits nontrivial $x$- and $n$-integrals if and
only if $d(t,t_1)$ is one of the kind:
\begin{enumerate}
\item[(1)] $d(t,t_1)=A(t_1-t)$, where $A(t_1-t)$ is given in
implicit form $A(t_1-t)=\frac{d}{d\theta}P(\theta)$,
$t_1-t=P(\theta)$, $P(\theta)$ is a quasi-polynomial on $\theta$,
\item[(2)] $d(t,t_1)=C_1(t_1^2-t^2)+C_2(t_1-t)$,
\item[(3)] $d(t,t_1)=\sqrt{C_3e^{2\alpha t_1}+C_4e^{\alpha(t_1+t)}+C_3e^{2\alpha t}},$
\item[(4)] $d(t,t_1)= C_5(e^{\alpha t_1}-e^{\alpha t})+C_6(e^{-\alpha
t_1}-e^{-\alpha t}),$
\end{enumerate}
\noindent where $\alpha\ne 0$, $C_i$, $1\leq i\leq 6$,  are
arbitrary constants. Moreover, some nontrivial $x$-integrals $F$ and
$n$-integrals $I$ in each of the cases are
\begin{enumerate}
\item[i)] $F=x-\int^{t_1-t}\frac{ds}{A(s)}$, $I=L(D_x)t_x$, where $L(D_x)$ is a differential operator
which annihilates $\frac{d}{d\theta}P(\theta)$ where $D_x\theta=1$.
 \item[ii)]
$F=\frac{(t_3-t_1)(t_2-t)}{(t_3-t_2)(t_1-t)}$, $I=t_x-C_1t^2-C_2t,$
\item[iii)]
$F=\int^{t_1-t}\frac{\mathrm{e}^{-\alpha s}ds}{\sqrt{C_3\mathrm{e}^{2\alpha
s} +C_4\mathrm{e}^{\alpha
s}+C_3}}-\int^{t_2-t_1}\frac{ds}{\sqrt{C_3\mathrm{e}^{2\alpha
s}+C_4\mathrm{e}^{\alpha s}+C_3}}$, $I=2t_{xx}-\alpha t_x^2-\alpha C_3
\mathrm{e}^{2\alpha t}$,
\item[iv)]
$F=\frac{(\mathrm{e}^{\alpha t}-\mathrm{e}^{\alpha
t_2})(\mathrm{e}^{\alpha t_1}-\mathrm{e}^{\alpha
t_3})}{(\mathrm{e}^{\alpha  t}-\mathrm{e}^{ \alpha
t_3})(\mathrm{e}^{\alpha t_1}-\mathrm{e}^{\alpha t_2})}$, $I=t_x-C_5
e^{\alpha t}-C_6 e^{-\alpha t}$.
\end{enumerate}
\end{thm}
Equation of the form $\tau_x=A(\tau)$, where $\tau=t_1-t$,  is
integrated in quadratures. But to get the final answer one should
evaluate the integral and then find the inverse function. In the
Darboux integrable case 1) the problem is effectively solved. The
general solution is given in an explicit form
\begin{equation}\label{case1solved}
t(n,x)=t(0,x)+\sum^{n-1}_{j=0} P(x+c_j),
\end{equation}
where $t(0,x)$ and $c_j$ are arbitrary functions of $x$ and $j$
respectively. Actually we have
$\tau_x=P_{\theta}(\theta)\theta_x=P_{\theta}(\theta)$, which
implies $\theta_x=1$, so that $\tau(n,x)=P(x+c_n)$. By solving the
equation $t(n+1,x)-t(n,x)=P(x+c_n)$ one gets the answer above.

The $x$-integrals in the cases 2) and 4) given in Theorem \ref{maintheorem} are given as double
relations of four points $t$, $t_1$, $t_2$, $t_3$ and respectively
points $e^t$, $e^{t_1}$, $e^{t_2}$, $e^{t_3}$. Due to the well known
theorem,  given four points  $z_1$, $z_2$, $z_3$, $z_4$
in the extended  complex plane $\bf{\hat{C}}$  can be converted to other
given four points  $w_1$, $w_2$, $w_3$, $w_4$ by one and the same
M\"{o}bius transformation
\begin{equation}\label{fractionallinear}
z=R(w):=\frac{a_{11}w+a_{12}}{a_{21}w+a_{22}}
\end{equation}
such that $z_j=R(w_j)$, where $j=1,2,3,4$, if and only if the points are connected by double relation
\begin{equation}\label{doublerelation}
\frac{z_{4}-z_{2}}{z_{4}-z_{3}}\frac{z_{3}-z_{1}}{z_{2}-z_{1}}= \frac{w_{4}-w_{2}}{w_{4}-w_{3}}\frac{w_{3}-w_{1}}{w_{2}-w_{1}}.
\end{equation}
Take a point $P(x)=(t(n+1,x),t(n+2,x),t(n+3,x),t(n+4,x))$ in
four-dimensional complex space $\bf{\hat{C}^{4}}$ compound by four
consecutive values of the variable $t$. The point $P(x)$ moves in
$\bf{\hat{C}^{4}}$ when $x$ ranges in the domain. Its trajectory coincides
with the orbit of point $P(x_0)$ under the action of transformations\footnote{The authors are
grateful to B.~G.~Konopelchenko, who paid their attention to this nice
geometric interpretation of the chains 2) and 4)}
$t(n+j,x)=R(t(n+j,x_0))$, $j=1,2,3,4$.

Studying the examples below we briefly discuss connection between discrete models and their continuum analogues.
The case 3) with $C_3=1$ and  $\alpha=1$ leads in the continuum limit to the equation
\begin{equation}\label{example1}
u_{xy}=e^u\sqrt{u_y^2+1}
\end{equation}
found earlier in \cite{Zhiber}. Indeed set $t(n,x)=u(y,x)$ and $C_4=-2+\epsilon^2,$ where $y=n\epsilon$. Then
substitute $\tau=\epsilon u_y+O(\epsilon^2)$ as $\epsilon\rightarrow 0$ into the equation $t_{1x}-t_x=e^t\sqrt{e^{2\tau}+C_4e^{\tau}+1}$ and evaluate the limit as $\epsilon\rightarrow 0$ to get (\ref{example1}). It is remarkable that equation (\ref{example1}) has the same integral ($y$-integral) $I=2u_{xx}-u_x^2-\mathrm{e}^{2u}$ as its discrete conterpart.

The chain $t_{1x}-t_x=(e^{t_1}-e^t)/2$ goes to the equation $u_{xy}=\frac{1}{2}e^u u_y$ in the continuum limit. Its $n$-integral $I=t_x-\frac{1}{2}e^t$  coincides with the corresponding $y$-integral of the continuum analogue. The Darboux integrable chain $t_{1x}-t_x=Ce^{(t_1+t)/2}$ (it comes from the case 3) for appropriate choice of the parameters) being  a discrete version of the  Liouville equation $u_{xy}=e^u$, also
has a common integral $I=2t_{xx}-t_x^2$ with its  continuum limit equation. Note that the chain defines the B\"{a}cklund transform for the Liouville equation.

The article is organized as follows. In the second section general results related to the Lie algebra
$L_n$ of equation (\ref{dhyp}) are given.  The third section is splitted into four subsections. Theorem  \ref{thm} from the Introduction gives a complete
list of equations (\ref{main}) admitting nontrivial $x$-integral. This list consists of four
different types of equations (\ref{main}).
In each  subsection  of Section 3  one of these four different types  from Theorem \ref{thm} is treated by imposing additional condition for an equation to possess nontrivial  $n$-integrals. The conclusion is provided in Section 4.

\section{General results}

Define a class \textbf{F} of locally analytic functions each of
which depends only on a finite number of dynamical variables. In
particular we assume that $f(t,t_1,t_x)\in \textbf{F}$. We will
consider vector fields given as infinite formal series of the form
\begin{equation}\label{formal}
Y=\sum_{k=0}^\infty y_k\frac{\partial}{\partial t_{[k]}}
\end{equation}
with coefficients $y_k\in \textbf{F}$. Introduce notions of
linearly dependent and independent sets of the vector fields
(\ref{formal}). Denote through $P_N$ the projection operator
acting according to the rule
\begin{equation}\label{projection}
P_N(Y)= \sum_{k=0}^{N} y_k\frac{\partial}{\partial t_{[k]}}.
\end{equation}
First we consider finite vector fields as
\begin{equation}\label{finitefield}
Z=\sum_{k=0}^{N}z_k\frac{\partial}{\partial t_{[k]}}.
\end{equation}
We say that a set of finite vector fields $Z_1$, $Z_2$, ..., $Z_m$
is linearly dependent in some open region \textbf{U}, if there is
a set of functions $\lambda_1,\,\lambda_2,\, ...,\lambda_m\in \textbf{F}$
defined on \textbf{U} such that the function
$|\lambda_1|^2+|\lambda_2|^2+...+|\lambda_m|^2$ does not vanish
identically and the condition
\begin{equation}\label{linear combination}
\lambda_1Z_1+\lambda_2Z_2+ ...+\lambda_mZ_m=0
\end{equation}
holds for each point of region \textbf{U}.

We call a set of the  vector fields $Z_1$, $Z_2$, ..., $Z_m$ of
the form (\ref{formal}) linearly dependent in the region
\textbf{U} if for each natural $N$ the following set of finite
vector fields $P_N(Z_1)$, $P_N(Z_2)$, ..., $P_N(Z_m)$ is linearly
dependent in this region. Otherwise we call the set $Z_1$, $Z_2$,
..., $Z_m$ linearly independent in \textbf{U}.

\noindent Now we give some properties of the characteristic Lie
algebra introduced in the Theorem \ref{thm1}. The proof of the
first two lemmas can be found in \cite{HabibullinPekcan}. However
for the reader's convenience we still give the proof of the second
Lemma.
\begin{lem}\label{linear}
If for some integer $N$ the operator $Y_{N+1}$ is a linear
combination of the operators $Y_i$ with $i\leq N$:
$Y_{N+1}=\alpha_1Y_1+\alpha_2 Y_2+...+\alpha_NY_N$, then for any
integer $j>N$, we have a similar expression
$Y_{j}=\beta_1Y_1+\beta_2 Y_2+...+\beta_NY_N$.
\end{lem}
\begin{lem}\label{lemma22} The following commutativity relations take place:
$ [Y_{0},X_1]=0$, $[Y_{0},Y_{1}]=0$ and $[X_1, DX_1D^{-1}]=0$.
\end{lem}

\noindent {\textbf{Proof.}} We have,
$$[Y_0,X_1]=\Big[\frac{d}{dt_1}, \frac{d}{dt_{-1}}\Big]=0,$$
$$[Y_0, Y_1]
=D^{-1}[DY_0D^{-1}, Y_0]D=D^{-1}\Big[\frac{d}{dt_2},
\frac{d}{dt_1}\Big]D=0, $$
$$[X_1, DX_1D^{-1}]=D[D^{-1}X_1D, X_1]D^{-1}=D[X_2, X_1]D^{-1}=0. \qquad \Box$$
Note that
\begin{equation}\label{Y_{k+1}=D^{-1}Y_kD}
Y_{k+1}=D^{-1}Y_kD, \qquad k\geq 2, \qquad D^{-1}Y_1D=X_1+Y_2\, .
\end{equation}
The next three statements turned out to be very useful for studying
the characteristic Lie algebra $L_n$.
\begin{lem}\label{Lemma6}\cite{HZhP}
If the Lie algebra generated by the vector fields $S_0=\sum_{j=-\infty}^\infty \frac{\partial }{\partial w_j}$
and \\$S_1=\sum_{j=-\infty}^\infty c(w_j)\frac{\partial }{\partial w_j}$ is of finite dimension then $c(w)$ is one of the
forms \\
(1) $c(w)=a_1+a_2e^{\lambda w}+a_3e^{-\lambda w}$, \\
(2) $c(w)=a_1+a_2w+a_3w^2$, where $\lambda\ne 0$, $a_1$, $a_2$ and  $a_3$ are some constants.
\end{lem}
\begin{lem}\label{lemma23} (1) Suppose that the vector field
$$
Y=\alpha(0)\frac{\partial }{\partial
t}+\alpha(1)\frac{\partial}{\partial
{t_x}}+\alpha(2)\frac{\partial }{\partial {t_{xx}}}+...,
$$
where $\alpha_x(0)=0$, solves the equation $[D_x,Y]=
\sum\limits_{k=-\infty,k\ne 0}^\infty \beta(k)\frac{\partial}{\partial t_k}$, then
$Y={\alpha(0)\frac{\partial}{\partial t}}$.\\
(2)  Suppose that the vector field
$$
Y=\alpha(1)\frac{\partial}{\partial {t_x}}+\alpha(2)\frac{\partial
}{\partial {t_{xx}}}+ \alpha(3)\frac{\partial }{\partial
{t_{xxx}}}+...
$$
 solves the equation $[D_x,Y]=h Y+
 \sum\limits_{k=-\infty,k\ne 0}^\infty \beta(k)\frac{\partial}{\partial t_k}$, where $h$ is a function of variables
 $t$, $t_x$, $t_{xx}$, $\ldots$, $t_{\pm 1}$, $t_{\pm 2}$, $\ldots$, then
$Y=0$.
\end{lem}

\begin{lem}\label{lemma24}  For any $m\geq 0$, we have
\begin{equation}\label{maincommrel}
[D_x,Y_m]=- \sum_{j=1}^mD^{-j}(Y_{m-j}(f))Y_j-
\sum\limits_{k=1}^\infty Y_m(D^{-(k-1)}g)\frac{\partial}{\partial
t_{-k}} -\sum\limits_{k=1}^\infty
Y_m(D^{k-1}f)\frac{\partial}{\partial t_{k}}\, .
\end{equation}
In particular,
\begin{equation}\label{DxY0}
[D_x, Y_0]=-\sum\limits_{k=1}^\infty Y_0(D^{k-1}f)\frac{\partial}{\partial t_{k}}\, .
\end{equation}

\begin{equation}\label{DxY1}
[D_x,Y_1]=-D^{-1}(Y_0(f))Y_1 -
\sum\limits_{k=1}^\infty Y_1(D^{-(k-1)}g)\frac{\partial}{\partial t_{-k}}
-\sum\limits_{k=1}^\infty Y_1(D^{k-1}f)\frac{\partial}{\partial t_{k}}\, .
\end{equation}

\end{lem}

\noindent Both  Lemmas \ref{lemma23} and \ref{lemma24}   easily  can be derived
from the following formula
\begin{eqnarray}\label{DxY}
[D_x,Y]&=&(\alpha_x(0)-\alpha(1)){\frac{\partial}{\partial
t}}-\sum\limits_{k=1}^\infty Y(D^{-(k-1)}g)\frac{\partial}{\partial
t_{-k}}\nonumber\\&-& \sum\limits_{k=1}^\infty
Y(D^{k-1}f)\frac{\partial}{\partial t_{k}}+\sum\limits_{k=1}^\infty
(\alpha_x(k)-\alpha(k+1))\frac{\partial }{\partial t_{[k]}}\,.
\end{eqnarray}

\noindent Suppose that the equation (\ref{dhyp}) admits a nontrivial
$n$-integral. Then, by Theorem \ref{thm1},  its characteristic Lie
algebra $L_n$ is of finite dimension. Linear space of the basic
vector fields $\{Y_k\}_1^{\infty}$ is also finite dimensional. We
have the following Theorem.
\begin{thm}
Dimension of $\mathrm{span}\{Y_k\}_1^{\infty}$ is finite and
equals, say $N$ if and only if the following system of equations
is consistent:
\begin{eqnarray}\label{systemD_x(lambda)}
D_x(\lambda_N)&=&\lambda_N(A_{N,N}-A_{N+1,N+1})-A_{N+1,N},\nonumber\\
D_x(\lambda_{N-1})&=&\lambda_{N-1}(A_{N-1,N-1}-A_{N+1,N+1})+\lambda_NA_{N,N-1}-A_{N+1,N-1},\nonumber\\
D_x(\lambda_{N-2})&=&\lambda_{N-2}(A_{N-2,N-2}-A_{N+1,N+1})+\lambda_{N-1}A_{N-1,N-2}
+\lambda_NA_{N,N-2}-A_{N+1,N-2},\nonumber\\
&\vdots&\nonumber\\
D_x(\lambda_2)&=&\lambda_2(A_{2,2}-A_{N+1,N+1})+\lambda_3A_{3,2}+...+\lambda_NA_{N,2}-A_{N+1,2},\nonumber\\
D_x(\lambda_1)&=&\lambda_1(A_{1,1}-A_{N+1,N+1})+\lambda_2A_{2,1}+\lambda_3A_{3,1}+...+\lambda_NA_{N,1}-A_{N+1,1}.\nonumber\\
0&=&\lambda_1A_{1,0}+\lambda_2A_{2,0}+\lambda_3A_{3,0}+...+\lambda_NA_{N,0}-A_{N+1,0}.
\end{eqnarray}
Here $A_{k,j}=D^{-j}(Y_{k-j}f)$.
\end{thm}
\noindent \textbf{Proof.} Suppose that the dimension of
$\mathrm{span}\{Y_k\}_1^{\infty}$ is finite, say N, then, by Lemma
\ref{linear}, $Y_1,...,Y_N$ form a basis in this linear space. So we
can find factors $\lambda_1,...,\lambda_N$ such that
\begin{equation}\label{YN+1=Y_1..Y}
Y_{N+1}=\lambda_1Y_1+\lambda_2Y_2+...+\lambda_NY_N.
\end{equation}
Take the commutator of both sides with $D_x$ and get by using the
main commutativity relation (\ref{maincommrel}) the following
equation,
\begin{eqnarray*}
-\sum_{j=0}^{N+1}A_{N+1,j}Y_j&=&D_x(\lambda_1)Y_1+D_x(\lambda_2)Y_2+...+D_x(\lambda_N)Y_N\\
&&-\Big(\lambda_1\sum_{j=0}^1A_{1,j}Y_j+\lambda_2\sum_{j=0}^2A_{2,j}Y_j+...+\lambda_N\sum_{j=0}^NA_{N,j}Y_j\Big).
\end{eqnarray*}
Now replace $Y_{N+1}$ at the left hand side by (\ref{YN+1=Y_1..Y}) and collect
coefficients of the independent vector fields to derive the system
given in the Theorem.\\

Suppose now, that the system (\ref{systemD_x(lambda)}) in the
Theorem has a solution. Let us prove that the vector field $Y_{N+1}$
is expressed in the form (\ref{YN+1=Y_1..Y}). Let
\begin{equation}
Z=Y_{N+1}-\lambda_1Y_1-\lambda_2Y_2-...-\lambda_NY_N.
\end{equation}
Let us find $[D_x,Z]$.
\begin{eqnarray*}
[D_x,Z]=[D_x,Y_{N+1}]-D_x(\lambda_1)Y_1-...-D_x(\lambda_N)Y_N
-\lambda_1[D_x,Y_1]-\lambda_2[D_x,Y_2]-...-\lambda_N[D_x,Y_N]
\\
=-\sum_{j=0}^{N+1}A_{N+1,j}Y_j-D_x(\lambda_1)Y_1-...-D_x(\lambda_N)Y_N
\\-\Big(\lambda_1\sum_{j=0}^1A_{1,j}Y_j+\lambda_2\sum_{j=0}^2A_{2,j}Y_j+...
+\lambda_N\sum_{j=0}^N A_{N,j}Y_j \Big)
+\sum\limits_{k=-\infty, k\ne 0}^\infty \beta(k) \frac{\partial}{\partial t_k}\, .
\end{eqnarray*}
Replace now $D_x(\lambda_1),...,D_x(\lambda_N)$ by means of the
system (\ref{systemD_x(lambda)}). After some simplifications one
gets
\begin{equation}\label{[D_x,Z]}
[D_x,Z]=-A_{N+1,N+1}Z+\sum\limits_{k=-\infty, k\ne 0}^\infty \beta(k) \frac{\partial}{\partial t_k}.
\end{equation}
By Lemma \ref{lemma23} we get $Z=0$. $\Box$\\

\noindent The proof of the next three results can be found in
\cite{TJM}.
\begin{lem}\label{lemma26} If the operator $Y_2=0$ then $[X_1,Y_1]=0$.
\end{lem}
The reverse statement to Lemma \ref{lemma26} is not true as the
equation $t_{1x}=t_x+e^t$ shows (see Lemma 3.4 below).

\begin{lem}\label{lemma27}
The operator $Y_2=0$ if and only if we have
\begin{equation}\label{Y2=0relation}
f_t+D^{-1}(f_{t_1})f_{t_x}=0.
\end{equation}
\end{lem}

\begin{cor}\label{cornint}
The dimension of the Lie algebra $L_n$ associated with
$n$-integral is equal to $2$ if and only if (\ref{Y2=0relation})
holds, or the same $Y_2=0$.
\end{cor}

\noindent Now let us introduce vector fields
\begin{equation}
\label{definitionC_n} C_1=[X_1,Y_1], \qquad C_k=[X_1, C_{k-1}],
\quad k\geq 2.
\end{equation}
It is easy to see that
\begin{equation}\label{coordinateC_n}
C_n=X_1^nD^{-1}\Big(Y_0(f)\Big)\frac{\partial}{\partial
t_x}+X_1^nD^{-1}\Big(Y_0D_x(f)\Big)\frac{\partial}{\partial
t_{xx}}+X_1^nD^{-1}\Big(Y_0D_x^2(f)\Big)\frac{\partial}{\partial
t_{xxx}}+...\,.
\end{equation}

\begin{lem}\label{Lemma2}
We have,
\begin{equation}\label{DxCn}
[D_x,C_n]=-g_{t_x}X_1^nD^{-1}Y_0(f)X_1-X_1^nD^{-1}Y_0(f)Y_1-\sum_{j=1}^nA_j^{(n)}C_j\,
,
\end{equation}
where
$$
A_j^{(n)}=X_1^{n-j}\left\{C(n,j-1)g_{t_{-1}}-C(n,j)\frac{g_t}{g_{t_x}}\right\},
\quad n\geq 1,\qquad C(n,k)=\frac{n!}{k!(n-k)!}.
$$
In particular,
$$
[D_x,
C_1]=-g_{t_x}X_1D^{-1}Y_0(f)X_1-X_1D^{-1}Y_0(f)Y_1-\left(g_{t_{-1}}-\frac{g_t}{g_{t_x}}\right)C_1\,
.
$$
\end{lem}

\noindent{\bf Proof}. We prove the Lemma by induction on  $n$.
Note that for any vector field
$$A=\beta(0)\frac{\partial }{\partial t}+\beta(1)\frac{\partial }{\partial t_x}+
\beta(2)\frac{\partial }{\partial t_{xx}}+\ldots \, ,
$$
acting on the set of functions $H$ depending on variavles $t_{-1}$, $t$, $t_{[k]}$, $k\in \NN$, formula (\ref{DxY})
becomes
\begin{eqnarray*}
&&[D_x, A]=-(\beta(0)g_t+\beta(1)g_{t_x})\frac{\partial }{\partial
t_{-1}}+ (\beta_x(0)-\beta(1))\frac{\partial }{\partial t}\\
&&+(\beta_x(1)-\beta(2))\frac{\partial }{\partial
t_x}+(\beta_x(2)-\beta(3))\frac{\partial }{\partial t_{xx}}+
(\beta_x(3)-\beta(4))\frac{\partial }{\partial t_{xxx}}+\
\end{eqnarray*}

Applying the last formula with $C_1$ instead of $A$, we have
\begin{eqnarray*}
[D_x,C_1]&=&-g_{t_x}X_1D^{-1}Y_0(f)X_1-X_1D^{-1}Y_0(f)\frac{\partial
}{\partial t}\\
&&+\sum_{k=1}^{\infty} \left\{D_xX_1D^{-1}Y_0D_x^{k-1}(f)
-X_1D^{-1}Y_0D_x^k(f) \right\}\frac{\partial}{\partial t_{[k]}} \, .
\end{eqnarray*}
Since
$$
[Y_0,D_x]G(t,t_1,t_x,t_{xx},
t_{xxx},\ldots)=f_{t_1}G_{t_1}=f_{t_1}Y_0G, \quad {\mbox{i.e.}}
\quad Y_0D_x=D_xY_0+f_{t_1}Y_0
$$
and
$$
[D_x, X_1]H(t_{-1},t, t_x, t_{xx}, t_{xxx},
\ldots)=-g_{t_{-1}}H_{t_{-1}}=-g_{t_{-1}}X_1H\, ,
$$
then
\begin{eqnarray*}
&&D_xX_1D^{-1}Y_0
D_x^{k-1}(f)-X_1D^{-1}Y_0D_x^k(f)=\left\{D_xX_1D^{-1}Y_0-X_1D^{-1}Y_0D_x\right\}D_x^{k-1}(f)\\
&&=\left\{D_xX_1D^{-1}Y_0-X_1D^{-1}\{ D_xY_0+f_{t_1}Y_0\}
\right\}D_x^{k-1}(f)\\
&&=[D_x,X_1]D^{-1}Y_0D_x^{k-1}(f)
X_1(D^{-1}Y_0(f))D^{-1}Y_0D_x^{k-1}(f)-D^{-1}(Y_0(f))X_1D^{-1}Y_0D_x^{k-1}(f)\\
&&=-g_{t_{-1}}X_1D^{-1}Y_0 D_x^{k-1}(f) -
X_1D^{-1}(Y_0(f))D^{-1}Y_0D_x^{k-1}(f)-D^{-1}(Y_0(f))X_1D^{-1}Y_0D_x^{k-1}(f).
\end{eqnarray*}
Therefore,
\begin{eqnarray*}
[D_x,
C_1]&=&-g_{t_x}X_1D^{-1}Y_0(f)X_1-X_1D^{-1}Y_0(f)\frac{\partial}{\partial
t}-\sum_{k=1}^\infty
X_1(D^{-1}Y_0(f))D^{-1}Y_0D_x^{k-1}(f)\frac{\partial }{\partial
t_{[k]}}\\
&&-g_{t_{-1}}\sum_{k=1}^\infty
X_1D^{-1}Y_0D_x^{k-1}(f)\frac{\partial }{\partial t_{[k]}}
-D^{-1}(Y_0(f))\sum_{k=1}^\infty
X_1D^{-1}Y_0D_x^{k-1}(f)\frac{\partial }{\partial t_{[k]}}\\
&=&-g_{t_x}X_1D^{-1}Y_0(f)X_1-X_1D^{-1}Y_0(f)Y_1-\left(g_{t_{-1}}+D^{-1}(f_{t_1})\right)C_1\,
.
\end{eqnarray*}
that proves the base of Mathematical induction. Assuming the
equation (\ref{DxCn}) is true for $n-1$, we have
\begin{eqnarray*}
[D_x, C_n]&=&[D_x, [X_1, C_{n-1}]]=-[X_1, [C_{n-1},
D_x]]-[C_{n-1},
[D_x, X_1]]\\
&=&[X_1, [D_x, C_{n-1}]]+[C_{n-1}, g_{t_{-1}}X_1]=[X_1, [D_x,
C_{n-1}]]+C_{n-1}(g_{t_{-1}})X_1-g_{t_{-1}}C_n\\
&=&[X_1,
-g_{t_x}X_1^{n-1}D^{-1}Y_0(f)X_1-X_1^{n-1}D^{-1}Y_0(f)Y_1-\sum_{j=1}^{n-1}A_j^{(n-1)}C_j]\\
&&+g_{t_{-1}t_x}X_1^{n-1}D^{-1}Y_0(f)X_1
-g_{t_{-1}}C_n\\
&=&-g_{t_{-1}t_x
}X_1^{n-1}D^{-1}Y_0(f)X_1-g_{t_x}X_1^nD^{-1}Y_0(f)X_1
-X_1^nD^{-1}Y_0(f)Y_1\\
&&-X_1^{n-1}D^{-1}Y_0(f) C_1
-\sum_{j=1}^{n-1}X_1(A_j^{(n-1)})C_j-\sum_{j=1}^{n-1}A_j^{(n-1)}C_{j+1}\\
&&+g_{t_{-1}t_x}X_1^{n-1}D^{-1}Y_0(f)X_1 -g_{t_{-1}}C_n\\
&=&-g_{t_x}X_1^nD^{-1}Y_0(f)X_1-X_1^nD^{-1}Y_0(f)Y_1-\{
A_{n-1}^{(n-1)}+g_{t_{-1}}\}C_n\\
&&-\{X_1^{n-1}D^{-1}Y_0(f)+X_1(A_1^{(n-1)}) \}C_1-
\sum_{j=2}^{n-1}\{X_1(A_j^{(n-1)})+A_{j-1}^{(n-1)}\}C_j\\
&=&-g_{t_x}X_1^nD^{-1}Y_0(f)X_1-X_1^nD^{-1}Y_0(f)Y_1-\sum_{j=1}^nA_j^{(n)}C_j\,
,
\end{eqnarray*}
where
\begin{eqnarray*}
A_1^{(n)}&=&X_1^{n-1}D^{-1}Y_0(f)+X_1(A_1^{(n-1)})\\&=&X_1^{n-1}\left\{-\frac{g_t}{g_{t_x}}\right\}+X_1X_1^{n-2}\left\{
C(n-1,0)g_{t_{-1}}-C(n-1,1)\frac{g_t}{g_{t_x}}\right\}\\
&=& X_1^{n-1}\{ C(n,0)g_{t_{-1}}-C(n, 1)\frac{g_t}{g_{t_x}}\}\, ;
\end{eqnarray*}
\begin{eqnarray*}
A_j^{(n)}=X_1(A_j^{(n-1)})+A_{j-1}^{(n-1)}&=&X_1X_1^{n-1-j}\left\{
C(n-1,j-1)g_{t_{-1}}-C(n-1,j)\frac{g_t}{g_{t_x}}\right\}\\
&&+X_1^{n-j}\left\{ C(n-1,j-2)g_{t_{-1}}-C(n-1,j-1)
\frac{g_t}{g_{t_x}}\right\}\\ &=&X_1^{n-j}\left\{
C(n,j-1)g_{t_{-1}}-C(n,j)\frac{g_t}{g_{t_x}}\right\}\, ;
\end{eqnarray*}
$$
A_n^n=A_{n-1}^{(n-1)}+g_{t_{-1}}=(n-1)g_{t_{-1}}-\frac{g_t}{g_{t_x}}+g_{t_{-1}}=ng_{t_{-1}}-\frac{g_t}{g_{t_x}}
$$
that finishes the proof of the Lemma. $\Box$

Assume  equation $t_{1x}=f(t,t_1,t_x)$ admits a nontrivial
$n$-integral.  Then we know that the dimension of Lie algebra
$L_n$ is at least $2$ by Corollary \ref{cornint}.

Consider case when the dimension of $L_n$ is at least 3 and
$C_1\ne 0$. Since linear space generated by vector fields $C_1$,
$C_2$, $C_3$, $\ldots$, is of finite dimension, then there exists
a natural number $N$ such that
$$
C_{N+1}=\mu_1C_1+\mu_2 C_2+\ldots +\mu_NC_N,
$$
and   $C_1$, $C_2$, $\ldots$, $C_N$ are linearly independent. By
Lemma \ref{Lemma2}    we have,
\begin{eqnarray*}
[D_x,C_{N+1}]&=&-g_{t_x}A_0^{(N+1)}X_1
-A_0^{(N+1)}Y_1-A_1^{(N+1)}C_1-\ldots - A_N^{(N+1)}C_n\\
&&-A_{N+1}^{(N+1)} \{ \mu_1C_1+\mu_2C_2+\ldots +\mu_N C_N\}   \, ,
\end{eqnarray*}
where $A_0^{(k)}=X_1^kD^{-1}Y_0(f)$. On the other hand,
$$
[D_x, C_{N+1}]=D_x(\mu_1)C_1+D_x(\mu_2)C_2+\ldots +D_x(\mu_N)C_N+
\mu_1(-g_{t_x}A_0^{(1)}X_1-A_0^{(1)}Y_1-A_1^{(1)}C_1)
$$
$$
+\mu_2(-g_{t_x}A_0^{(2)}X_1-A_0^{(2)}Y_1-A_1^{(2)}C_1-A_2^{(2)}C_2)
+\ldots +\mu_N
(-g_{t_x}A_0^{(N)}X_1-A_0^{(N)}Y_1-A_1^{(N)}C_1-\ldots
-A_N^{(N)}C_N)  \, .
$$
Linear independence of $X_1$, $Y_1$, $C_1$, $C_2$, $\ldots$, $C_N$
allows us to compare coefficients before $X_1$, $C_k$, $1\leq
k\leq N$ in the last two presentations for $[D_x,C_{N+1}]$ . We
have,
\begin{equation}\label{systemnintegral}\begin{array}{l}
-A_0^{(N+1)}=-\mu_1 A_0^{(1)}-\mu_2A_0^{(2)}-\ldots -\mu_N
A_0^{(N)}\,    ,
\\
-A_1^{(N+1)}-\mu_1A_{N+1}^{(N+1)}=  -\mu_1
A_1^{(1)}-\mu_2A_1^{(2)}-\ldots -\mu_N A_1^{(N)}+D_x(\mu_1)\,  ,
\\
-A_k^{(N+1)}-\mu_kA_{N+1}^{(N+1)}= -\{\sum_{j=k}^N
\mu_jA_k^{(j)}\}+D_x(\mu_k)\, , \quad 2\leq k\leq  N-3
 ,\\
 -A_{N-2}^{(N+1)}-\mu_{N-2}A_{N+1}^{(N+1)}=  -\mu_{N-2} A_{N-2}^{(N-2)}-\mu_{N-1}A_{N-2}^{(N-1)}-\mu_N A_{N-2}^{(N)}
 +D_x(\mu_{N-2})\,  ,          \\
 -A_{N-1}^{(N+1)}-\mu_{N-1}A_{N+1}^{(N+1)}=  -\mu_{N-1} A_{N-1}^{(N-1)}-\mu_N A_{N-1}^{(N)}
+D_x(\mu_{N-1})\, ,  \\
    -A_{N}^{(N+1)}-\mu_{N}A_{N+1}^{(N+1)}=  -\mu_{N} A_{N}^{(N)}+D_x(\mu_{N})\,
    .
 \end{array}
 \end{equation}
 Thus we have proved the following Theorem.
 \begin{thm} Consistency of the system (\ref{systemnintegral})
 is necessary for existence of a nontrivial n-integral to the chain (\ref{dhyp}).
 \end{thm}

One can specify the system. Since
  $$
  \begin{array}{l}
  A_N^{(N+1)}=X_1\left\{C(N+1,N-1)g_{t_{-1}}-C(N+1,N)\frac{g_t}{g_{t_x}}\right\}=\frac{(N+1)N}{2}g_{t_{-1}t_{-1}}
  -(N+1)\frac{g_{tt_{-1}}g_{t_x}-g_tg_{t_x t_{-1}}  }{g_{t_x}^2}\, ,\\
  A_{N+1}^{(N+1)}=
  \left\{C(N+1,N)g_{t_{-1}}-C(N+1,N+1)\frac{g_t}{g_{t_x}}\right\}=(N+1)g_{t_{-1}}-\frac{g_t}{g_{t_x}}\, ,\\
  A_N^{(N)}=  \left\{C(N,N-1)g_{t_{-1}}-C(N,N)\frac{g_t}{g_{t_x}}\right\}
  =  Ng_{t_{-1}}-\frac{g_t}{g_{t_x}}\,   ,
  \end{array}
  $$
 the last equation of (\ref{systemnintegral}) becomes
\begin{eqnarray*}
&& \left\{ \frac{(N+1)N}{2}g_{t_{-1}t_{-1}}
-(N+1)\frac{g_{tt_{-1}}g_{t_x}-g_tg_{t_x t_{-1}}  }{g_{t_x}^2}
\right \}     +\mu_N    \left \{
(N+1)g_{t_{-1}}-\frac{g_t}{g_{t_x}}\right\}=\\
&&=\mu_N\left\{ Ng_{t_{-1}}-\frac{g_t}{g_{t_x}}\right\}
-D_x(\mu_N) \, ,
\end{eqnarray*}
 that can be rewritten as
 \begin{equation}
 \label{nintegralEq1}
 \frac{(N+1)N}{2} g_{t_{-1}t_{-1}}-(N+1) \frac{g_{tt_{-1}}g_{t_x}-g_tg_{t_x t_{-1}}  }{g_{t_x}^2}
 +\mu_Ng_{t_{-1}}=-D_x(\mu_N)   \, .
 \end{equation}
   If $C_1=0$ then, by Lemma \ref{Lemma2},
   \begin{equation}\label{C_1=0}
  0=X_1D^{-1}Y_0(f)=\frac{\partial }{\partial t_{-1}}D^{-1}(f_{t_1})
  \, .
   \end{equation}

\section{ Proof of  Theorem \ref{maintheorem}}

\subsection{Case 1) $t_{1x}=t_x+A(t_1-t)$}

\noindent Introduce $\tau=t_1-t$ and rewrite the equation as
$\tau_x=A(\tau)$. Study the question when this equation admits a
nontrivial $n$-integral or the same when the corresponding Lie
algebra $L_n$ is of finite dimension. Since
$$Y_0f=A'(\tau){\tau}_{t_1}=D_{\tau}A(\tau),$$
$$Y_0f_x=A''(\tau)A(\tau)+A'(\tau)A'(\tau)=D_{\tau}A(\tau)D_{\tau}A(\tau),$$
and $Y_0D_x^kf=(D_{\tau}A(\tau))^{k+1}$, we can write $Y_1$ as
\begin{eqnarray}\label{case1Y_1}
Y_1=\frac{\partial}{\partial
t}+\sum_{k=1}^{\infty}D^{-1}(D_{\tau}A(\tau))^k\frac{\partial}{\partial
D_x^kt}.
\end{eqnarray}
\noindent Now let us introduce new variables: $\tau_{+}=t,
\tau=t_1-t,  \tau_{-1}=t-t_{-1}, \tau_j=t_{j+1}-t_j$. Since
\begin{eqnarray*}
\frac{\partial}{\partial t}&=&\frac{\partial}{\partial \tau_{+}}-
\frac{\partial}{\partial \tau}+\frac{\partial}{\partial \tau_{-1}},\\
\end{eqnarray*}
then  the expression (\ref{case1Y_1}) for $Y_1$ becomes
\begin{equation}\label{case1newY_1}
Y_1=\frac{\partial}{\partial \tau_{+}}-\frac{\partial}{\partial
\tau}+\frac{\partial}{\partial
\tau_{-1}}+\sum_{k=1}^{\infty}D^{-1}(D_{\tau}A(\tau))^k\frac{\partial}{\partial
D_x^k\tau_{+}}.
\end{equation}
\noindent One can ignore the term containing
$\frac{\partial}{\partial \tau}$ since coefficients in the vector fields used below do not
depend on $\tau$.\\

\noindent Multiply $Y_1$ by $A(\tau_{-1})$,
\begin{equation}\label{case1AY_1}
A(\tau_{-1})Y_1=A(\tau_{-1})\frac{\partial}{\partial
\tau_{+}}+A(\tau_{-1})\frac{\partial}{\partial
\tau_{-1}}+\sum_{k=1}^{\infty}A(\tau_{-1})D^{-1}(D_{\tau}A(\tau))^k\frac{\partial}{\partial
D_x^k\tau_{+}}.
\end{equation}

\noindent Introduce \begin{equation}\label{p}
p(\theta)=A(\tau_{-1}(\theta)), \qquad {\mbox{where}} \qquad
d\theta=\frac{d\tau_{-1}}{A(\tau_{-1})}.\end{equation} The equation
(\ref{case1AY_1}) becomes
\begin{equation}\label{case1eqn3}
A(\tau_{-1})Y_1=p(\theta)\frac{\partial}{\partial
\tau_{+}}+\frac{\partial}{\partial
\theta}+\sum_{k=1}^{\infty}D_x^k(p(\theta))\frac{\partial}{\partial
D_x^k\tau_{+}}.
\end{equation}
\noindent Now instead of $X_1=\frac{\partial}{\partial t_{-1}}$,
define
$$\tilde{X}_1=A(\tau_{-1})X_1=-A(\tau_{-1})\frac{\partial}{\partial
\tau_{-1}}+A(\tau_{-1})\frac{\partial}{\partial \tau_{-2}}.$$ It is
indeed with new variables
\begin{equation}\label{tilde{X}1}
\tilde{X}_1=-\frac{\partial}{\partial
\theta}+\frac{p(\theta)}{p(\theta_{-1})}\frac{\partial}{\partial
\theta_{-1}}.
\end{equation}
\noindent Note that
$[D_x,\tilde{X}_1]=D_x\Big(\frac{p(\theta)}{p(\theta_{-1})}\Big)W_1$,
where $W_1=\frac{\partial}{\partial \theta_{-1}}$.  Since $[D_x, X_1]=-X_1(g)X_1-X_1(g_{-1})X_2$, then
$[D_x, \tilde{X}_1]\in L_n$. Therefore, we have
two possibilities;
\begin{enumerate}
\item[i)]$D_x\Big(\frac{p(\theta)}{p(\theta_{-1})}\Big)=0$, or\\
\item[ii)]$W_1\in L_n$.
\end{enumerate}
\noindent First let us consider  case i). We have
$$D_x\Big(\frac{p(\theta)}{p(\theta_{-1})}\Big)=\frac{p'(\theta)p(\theta_{-1})
-p(\theta)p'(\theta_{-1})}{p^2(\theta_{-1})}=0.$$ Solving this differential equation we get $p(\theta)=A(\tau_{-1}(\theta))=\mu
e^{\lambda \theta}$. Since
$\frac{d\theta}{d\tau_{-1}}=\frac{1}{A(\tau_{-1})}$, we have
$A(\tau)=\lambda \tau+c$.\\

\noindent Now concentrate on case ii). Since
$D_x\Big(\frac{p(\theta)}{p(\theta_{-1})}\Big)W_1\in L_n$, then
$W_1\in L_n$ and, due to (\ref{tilde{X}1}),
$W=\frac{\partial}{\partial \theta}\in L_n$.

\begin{lem}\label{cor1}
If equation $\tau_x=A(\tau)$ admits a nontrivial $n$-integral then
function $p(\theta)$, defined by (\ref{p}), is a quasi-polynomial.
\end{lem}

\noindent \textbf{Proof.} Instead of $Y_1, X_1$, take the pair of
the operators $W=\frac{\partial}{\partial \theta}$ and
\begin{equation}\label{Z}
Z=A(\tau_{-1})Y_1-W=p(\theta)\frac{\partial}{\partial
\tau_{+}}+D_xp(\theta)\frac{\partial}{\partial
\tau_{+x}}+D_x^2(p(\theta))\frac{\partial}{\partial
\tau_{+xx}}+...\,.
\end{equation}
Construct a sequence of the operators
\begin{equation}\label{seqC_j}
C_1=[W,Z],\quad C_2=[W,C_1],\quad C_k=[W,C_{k-1}],\quad k\geq 2.
\end{equation}
Since algebra $L_n$ is of finite dimension then there exists
number $N$ such that
\begin{equation}\label{case1eqn6}
C_{N+1}=\mu_0Z+\mu_1C_1+...+\mu_NC_N,
\end{equation}
and vector fields $Z$, $C_1$, $\ldots$, $C_N$ are linearly
independent.\\
Direct calculations show that $[D_x,W]=[D_x,Z]=0$. Therefore,  we
have $[D_x,C_j]=0$ for all $j$. It follows from (\ref{case1eqn6})
that
$$0=D_x(\mu_0)Z+D_x(\mu_1)C_1+...+D_x(\mu_N)C_N,$$
which implies $D_x(\mu_j)=0$. Clearly $\mu_j=\mu_j(\theta)$ and
$D_x(\mu_j)=\mu_j'(\theta)=0$. Hence $\mu_j$ is constant for all
$j\geq 0$.\\
Look at the coefficients of $\frac{\partial}{\partial \tau_{+}}$ in
(\ref{case1eqn6}) and get
\begin{equation}\label{case1eqn6'}
\mu_0p(\theta)+\mu_1p'(\theta)+...+\mu_Np^{(N)}(\theta)=p^{(N+1)}(\theta).
\end{equation}
This means $p(\theta)$ is a quasi-polynomial, i.e. it takes the
form
\begin{equation}\label{quasipolyp}
p(\theta)=\sum_{j=1}^sq_j(\theta)e^{\lambda_j\theta}.
\end{equation}
$\Box$\\

\begin{lem}
Let $p(\theta)$ is an arbitrary quasi-polynomial solving a
differential equation of the form (\ref{case1eqn6'}) and which does
not solve any equation of this form of less order. Then the equation
$t_{1x}=t_x+A(t_1-t)$ with $A$ found from the conditions
\begin{eqnarray*}
A(\tau_{-1})=p(\theta),\\
\tau_{-1}=\int_0^{\theta}p(\tilde{\theta})d\tilde{\theta}
\end{eqnarray*}
admits a nontrivial n-integral.
\end{lem}

\noindent{\textbf{Proof.}} Introduce
$$L(D_x)=D_x^{N+1}-\mu_ND_x^N-\mu_{N-1}D_x^{N-1}-...-\mu_1D_x-\mu_0.$$
Equation (\ref{case1eqn6'}) can be rewritten as
$L(D_x)p(\theta)=0$. However $L(D_x)p(\theta)=L(D_x)A(\tau_{-1})$.
Since $L(D_x)t_{1x}=L(D_x)t_x+L(D_x)A(\tau)$ and
$L(D_x)A(\tau)=0$, we have $L(D_x)t_{1x}=L(D_x)t_x$. But
$L(D_x)t_{1x}=DL(D_x)t_x$, therefore $DL(D_x)t_x=L(D_x)t_x$.
Denote $L(D_x)t_x=I$ so we have $DI=I$. Hence $L(D_x)t_x$ is an
$n$-integral.$\Box$

Therefore the condition (\ref{quasipolyp}) is necessary and
sufficient for our equation to have nontrivial $n$-integral.

\noindent \textbf{Example 1.} Take
$p(\theta)=\frac{1}{2}e^{\theta}+\frac{1}{2}e^{-\theta}=\cosh\theta$,
then
\begin{eqnarray*}
A(\tau_{-1})&=&\cosh\theta\\
\tau_{-1}&=&\sinh\theta+c,
\end{eqnarray*}
or $A(\tau_{-1})^2-(\tau_{-1}-c)^2=1$ which gives
$A(\tau_{-1})=\sqrt{1+(\tau_{-1}-c)^2}$. So
$t_{1x}=t_x+\sqrt{1+(t_1-t-c)^2}$, where $c$ is arbitrary constant,
is Darboux integrable. Moreover, its general solution is given by $t(n,x)=G(x)+nc
+\sum_{k=0}^{n-1}\sinh (x+c_k)$, where $G(x)$ is arbitrary function
depending on $x$, and  $c_k$ are arbitrary  constants.

\subsection{Case 2) $t_{1x}=t_x+c_1(t_1-t)t+c_2(t_1-t)^2+c_3(t_1-t)$}

\begin{lem}\label{lemma}
If equation $t_{1x}=t_x+d(t,t_1)=t_x+c_1(t_1-t)t+c_2(t_1-t)^2+c_3(t_1-t)$ admits a nontrivial $n$-integral,
then there exists a natural number $k$ such that
\begin{equation}\label{c_1andc_2}
kc_1-(k+1)c_2=0\, .
\end{equation}
\end{lem}
{\textbf{Proof}.} Introduce vector fields $T_1=[X_1,Y_1]$,
$T_n=[X_1,T_{n-1}]$, $n\geq 2$. Direct calculations show that
$$
[D_x,T_1]=(-c_1+2c_2)X_1+(-c_1+2c_2)Y_1+(d_{t_{-1}}(t_{-1},t)-d_{t}(t_{-1},t))T_1,
$$
\begin{equation}\label{[D_x,T_n]}
[D_x,T_n]=-A_{n-1}^{(n)}T_{n-1}-A_n^{(n)}T_n,
\end{equation}
where
$$
A_j^{(n)}=X_1^{n-j}\{-C(n,j-1)d_{t_{-1}}(t_{-1},t)+C(n,j)d_{t}(t_{-1},t)\}, \qquad C(n,k)=\frac{n!}{k!(n-k)!}\, .
$$
Due to finiteness of algebra $L_n$, there exists  natural number $M$
such that
$$
T_{M+1}=\mu_1T_1+\mu_2T_2+\ldots +\mu_{M}T_M,
$$
and $T_1$, $T_2$, $\ldots$, $T_M$ are linearly independent.
We have,
$$
[D_x,T_{M+1}]=[D_x,\mu_1T_1+\mu_2T_2+\ldots +\mu_{M}T_M],
$$
that can be rewritten by (\ref{[D_x,T_n]}) in the following form:
\begin{eqnarray*}
&&-A_{M}^{(M+1)}T_M-A_{M+1}^{(M+1)}\{\mu_MT_M+\mu_{M-1}T_{M-1}+...+\mu_{1}T_1\}=D_x(\mu_1)T_1-\mu_1(c_1-2c_2)X_1\\
&& \hspace{44mm}-\mu_1(c_1-2c_2)Y_1-\mu_1A_1^{(1)}T_1+
D_x(\mu_2)T_2-\mu_2A_1^{(2)}T_1-\mu_2A_2^{(2)}T_2\\&&\hspace{70mm}+...+D_x(\mu_M)T_M-\mu_MA_{M-1}^{(M)}T_{M-1}-\mu_MA_{M}^{(M)}T_M.
\end{eqnarray*}
Compare coefficients before the operators. The coefficient before
$X_1$ and $Y_1$ gives $-\mu_1(c_1-2c_2)=0$. In this case we have
two choices: $\mu_1=0$ or $c_1-2c_2=0$. The second one gives (\ref{c_1andc_2})
with $k=1$. If $c_1-2c_2\ne 0$, then
 $\mu_1=0$. Using this, from the
coefficient of $T_1$ we get $-\mu_2A_1^{(2)}=0$. Again, we have that either
$\mu_2=0$, or $A_{1}^{(2)}=0$. If $A_{1}^{(2)}=0$ we stop, if not
then $\mu_2=0$ and we continue to compare the coefficients. Using
$\mu_1=\mu_2=0$, the coefficient before $T_2$ gives
$-\mu_3A_2^{(3)}=0$ which means $\mu_3=0$ or $A_2^{(3)}=0$. Same as
before: if $A_2^{(3)}=0$, we stop, if not then $\mu_3=0$ and we
continue to the procedure.

If $\mu_1=...=\mu_M=0$ then $T_{M+1}=0$ and
$[D_x,T_{M+1}]=0=-A_{M}^{(M+1)}T_M-A_{M+1}^{(M+1)}T_{M+1}=-A_M^{M+1}T_M$.
Since $T_1$, $\ldots$, $T_M$ are linearly independent then
$T_M\neq 0$ and therefore $A_M^{(M+1)}=0$. It follows
$A_{k-1}^{(k)}=0$ for some $k=1,2,...,M+1$. Evaluate
$A_{k-1}^{(k)}$:
\begin{eqnarray*}
A_{k-1}^{(k)}&=&-C(k,k-2)d_{t_{-1}t_{-1}}(t_{-1},t)+C(k,k-1)d_{tt_{-1}}(t_{-1},t)\\
&=&-k(k-1)(c_2-c_1)+k(c_1-2c_2)=k\{kc_1-(k+1)c_2\}.
\end{eqnarray*}$
\Box$\\

\noindent Let us rewrite the equation in case $2)$ as
$$\tau_x=c_1\tau t+c_2\tau^2+c_3\tau,$$
where $\tau=t_1-t$. We have two important
relations:\\

\noindent $1)$ $Y_0f=D_x\ln H$, where
\begin{equation}\label{defH}
H=\frac{\tau\theta^{1/\epsilon}}{(\theta+\epsilon)^{1/\epsilon}},
\qquad \theta=\frac{\tau_1}{\tau}, \qquad
\epsilon=\frac{c_1}{c_2}-1\, .
\end{equation}

\noindent $2)$ $Y_1 f=D_x\ln R H_{-1}$, where $$H_{-1}=D^{-1}H,
\qquad R=\frac{\theta}{\tau(\theta+\epsilon)}, \qquad {\mbox{ when}}
\qquad\epsilon\ne 0. $$(The case $\epsilon=0$, i.e $c_1=c_2$, is not
realized due to Lemma \ref{lemma}.
 The case $c_2=0$, due to Lemma \ref{lemma}, leads to $c_1=0$, and the equation becomes
 $t_{1x}=t_x+c_3(t_1-t)$ with an $n$-integral $I=t_x-c_3t$.)\\

\noindent These two relations allow us to simplify the basis
operators $Y_0,Y_1,X_1$. Really, we take
$$\tilde{Y}_1=H_{-1}Y_1,\quad \tilde{Y}_0=HY_0,$$
and get $[D_x,\tilde{Y}_0]=0$ and $[D_x,\tilde{Y_1}]=\Lambda
\tilde{Y}_0$, where $\Lambda=-\frac{H_{-1}}{H}D_x\ln(RH_{-1})$.\\

\noindent First we will restrict the set of the variables as
follows: $t_1,t,t_{-1},t_x,t_{xx},...$ and change the variables $t^{+}=t$, $\tau_{-1}=t-t_{-1}$
keeping the other variables unchanged. Then some of the differentiations will change
$$\frac{\partial}{\partial t}=\frac{\partial}{\partial t^{+}}+\frac{\partial}{\partial \tau_{-1}}
,\quad \frac{\partial}{\partial t_{-1}}=-\frac{\partial}{\partial \tau_{-1}}.$$
So we have $X_1=-\frac{\partial}{\partial \tau_{-1}}=-\hat{X}_1$ and
$$\tilde{Y}_1=H_{-1}\Big(\frac{\partial}{\partial t^{+}}+\frac{\partial}{\partial \tau_{-1}}\Big)+\sum\limits_{k=1}^\infty H_{-1}D^{-1}(Y_0D_x^{k-1}f)\frac{\partial}{\partial t_{[k]}}\,.$$
Since $[D_x,\hat{X}_1]=D_x(\ln R_{-1})\hat{X}_1$, one can introduce $\tilde{X}_1=\frac{1}{R_{-1}}\hat{X}_1$ and get $[D_x,\tilde{X}_1]=0$. Here
$R_{-1}=D^{-1}R$.

\noindent Introduce vector fields $C_2=[\tilde{X}_1,\tilde{Y}_1]$,
$C_3=[\tilde{X}_1,C_2]$, $C_k=[\tilde{X}_1,C_{k-1}]$, $k\geq 3$. We have,
$$[D_x,C_{j+1}]=\tilde{X}_1^j(\Lambda)\tilde{Y}_0,\quad j\geq 1.$$

\noindent Since the algebra $L_n$ is of finite dimension then
there is a number $N$ such that
\begin{equation}\label{star}
C_{N+1}=\mu_NC_N+...+\mu_2C_2+\mu_1\tilde{Y}_1,
\end{equation}
where $\tilde{Y}_1$, $C_1$, $C_2$, $\ldots$ are linearly
independent.\\

\noindent Applying the commutator with $D_x$ one gets
$D_x(\mu_j)=0$ for $j=1,...,N$ and
\begin{equation}\label{2star}
(\tilde{X}_1^N-\mu_N\tilde{X}_1^{N-1}-...-\mu_1)\Lambda=0.
\end{equation}
All the operators in our sequence have coefficients depending on
$\tau,\tau_{-1},t$. So do $\mu_j=\mu_j(\tau,\tau_{-1},t)$. But the
relation $D_x\mu_j(\tau,\tau_{-1},t)=0$ shows that
$\frac{\partial\mu_j}{\partial t}=0$ i.e.
$\mu_j=\mu_j(\tau,\tau_{-1})$. Since the minimal $x$-integral
 for an equation in case 2) depends on variables $t$, $t_1$, $t_2$, $t_3$, the relation
$D_x(\mu_j)=0$ implies that $\mu_j$ is constant for all $j$.\\

\noindent Introduce new variables $\tilde{t}_1$, $\tilde{t}$, $\eta$ as
$$\tilde{t}_1=t_1, \qquad \tilde{t}=t^{+},$$
\begin{equation}\label{defnu}
\eta=\ln\Big(\frac{\tau_{-1}}{\tau_{-1}+\frac{1}{\epsilon}(t_1-t^+)}\Big)\quad;\quad
{\mbox{or the same}}\qquad
\tau_{-1}=\frac{\tau}{\epsilon}\Big(\frac{e^{\eta}}{1-e^{\eta}}\Big).\end{equation}
Then
\begin{eqnarray*}
\frac{\partial}{\partial\tau_{-1}}&=&\frac{\partial\eta}{\partial\tau_{-1}}\frac{\partial}{\partial\eta},\\
\frac{\partial}{\partial t^{+}}&=&\frac{\partial}{\partial\tilde{t}}+\frac{\partial\eta}{\partial t^+}\frac{\partial}{\partial\eta},\\
\frac{\partial}{\partial
t_1}&=&\frac{\partial}{\partial\tilde{t}_1}+\frac{\partial\eta}{\partial
t_1}\frac{\partial}{\partial\eta}.
\end{eqnarray*}
In these new variables $\tilde{X}_1$ takes the form

$$\tilde{X}_1=\frac{\tau_{-1}(\theta_{-1}+\epsilon)}{\theta_{-1}}\frac{\partial}{\partial\tau_{-1}}=\frac{\partial}{\partial\eta}$$
and equation (\ref{2star}) becomes

\begin{equation}\label{new2star}
\Big(\frac{d^N}{d\eta^N}-\mu_N\frac{d^{N-1}}{d\eta^{N-1}}-...-\mu_1 \Big)\Lambda=0,
\end{equation}
where
\begin{equation}\label{Lambda}
\Lambda=-\frac{H_{-1}}{H}(\tau_x\ln R+D_x\ln H_{-1})
=-\frac{H_{-1}}{H}\Big(\frac{\partial f}{\partial
t}+D^{-1}\frac{\partial f}{\partial t_1}\Big)\\
=-\frac{H_{-1}}{H}(c_1-2c_2) (\tau-\tau_{-1}).
\end{equation}
Let us show that  $c_1-2c_2=0$. Assume contrary. It follows from
(\ref{new2star}) and (\ref{Lambda}) that both functions $H_{-1}$ and
$\tau_{-1}H_{-1}$ should solve the linear differential equation with
constant coefficients:
$$\Big(\frac{d^N}{d\eta^N}-\mu_N\frac{d^{N-1}}{d\eta^{N-1}}-...-\mu_1 \Big)y(\eta)=0.$$
Therefore, both functions $H_{-1}$ and $\tau_{-1}H_{-1}$ must be
quasi-polynomials in $\eta$.

\noindent Due to (\ref{defH}) and (\ref{defnu}), we have
$$H_{-1}=\frac{\tau}{\epsilon}
e^{\eta}(1-e^{\eta})^{\frac{1}{\epsilon}-1}$$ and
$$\tau_{-1}H_{-1}=\frac{\tau^2}{\epsilon^2}
e^{2\eta}(1-e^{\eta})^{\frac{1}{\epsilon}-2}\, .$$ To be
quasi-polynomials in $\eta$ it is necessary that
$\epsilon=\frac{1}{m}$
for some natural  $m\geq 2$.\\
Rewrite our vector fields $\tilde{X}_1,\tilde{Y}_1$ in the new
variables;
\begin{eqnarray*}
\tilde{X}_1&=&\frac{\partial}{\partial\eta},\\
\tilde{Y}_1&=&H_{-1}\frac{\partial}{\partial\tilde{t}}+
H_{-1}\Big(\frac{\partial\eta}{\partial
t^{+}}+\frac{\partial\eta}{\partial\tau_{-1}}
\Big)\frac{\partial}{\partial\eta}+...\,.
\end{eqnarray*}
Study the projection on the direction
$\frac{\partial}{\partial\eta}$.\\
The operators $\tilde{X}_1=\frac{\partial}{\partial \eta}$ and
$H_{-1}\Big(\frac{\partial\eta}{\partial
t^{+}}+\frac{\partial\eta}{\partial\tau_{-1}}\Big)\frac{\partial}{\partial\eta}$
generate a finite dimensional Lie algebra over the field of
constants. Due to  Lemma \ref{Lemma6}  in this case the coefficient
$H_{-1}\frac{\partial\eta}{\partial t}$ should be of one of the
forms
\begin{equation}\label{3star}
\tilde{c}_1e^{\tilde{\alpha}\eta}+\tilde{c}_2e^{-\tilde{\alpha}\eta}+\tilde{c}_3
\quad \mathrm{or} \quad
\tilde{c}_1\eta^2+\tilde{c}_2\eta+\tilde{c}_3,
\end{equation}
but we have
$$H_{-1}\Big(\frac{\partial\eta}{\partial t^{+}}+\frac{\partial\eta}{\partial\tau_{-1}}\Big)= \Big(1+\Big(\frac{1}{\epsilon}-1\Big)e^{\eta}\Big)(1-e^{\eta})^{\frac{1}{\epsilon}},$$
with $\frac{1}{\epsilon}=m\geq 2$ and it is never of the form (\ref{3star}). This contradiction
shows that $c_1-2c_2=0$. $\Box$\\

\subsection{Case 3) $t_{1x}=t_x+A(t_1-t)e^{\alpha t}$}

\noindent Introduce $\tau=t_1-t$ and rewrite the equation as
$\tau_x=A(\tau)e^{\alpha t}$. Study the question when the equation
admits a nontrivial $n$-integral or the same when the
corresponding
Lie algebra $L_n$ is of finite dimension.\\

\noindent Instead of the vector fields
$Y_0=\frac{\partial}{\partial t_1}$ and
$Y_1=\frac{\partial}{\partial t}+D^{-1}\Big(\frac{\partial
f}{\partial t_1}\Big)\frac{\partial}{\partial t_x}
+D^{-1}\Big(\frac{\partial f_x}{\partial
t_1}\Big)\frac{\partial}{\partial t_{xx}}+...$, we will use the
vector fields $\tilde{Y}_0=A(\tau)Y_0$ and
$\tilde{Y}_1=A(\tau_{-1})Y_1$. They are more convenient since they
satisfy more simple relations:
$$[D_x,\tilde{Y}_0]=0,\quad [D_x,\tilde{Y}_1]=\lambda_1\tilde{Y}_0$$
as operators acting on the enlarged set $t_1,t,t_{-1},t_{-2},...$
; $t_x,t_{xx},t_{xxx},...$\,. Here the coefficient $\lambda_1$ is
$$\lambda_1=\frac{A(\tau_{-1})}{A(\tau)}\Big(A'(\tau)-\alpha A(\tau)-A'(\tau_{-1})e^{-\alpha \tau_{-1}}\Big)e^{\alpha t}.$$

\noindent Since the equation is represented as
$\tau_x=A(\tau)e^{\alpha t}$ it is reasonable to introduce new
variables as $\tau_{+}=t, \tau_{-1}=t-t_{-1},
\tau_{-2}=t_{-1}-t_{-2}$, such that $$ \frac{\partial}{\partial
t}=\frac{\partial}{\partial \tau_{+}}+\frac{\partial}{\partial
\tau_{-1}},\quad \frac{\partial}{\partial
t_{-1}}=-\frac{\partial}{\partial
\tau_{-1}}+\frac{\partial}{\partial \tau_{-2}},\quad
\frac{\partial}{\partial t_{-2}}=-\frac{\partial}{\partial
\tau_{-2}}.$$

\noindent Instead of the operators $X_1=\frac{\partial}{\partial
t_{-1}}$ and $X_2=\frac{\partial}{\partial t_{-2}}$ use new ones
$\tilde{X}_1=A(\tau_{-1})e^{-\alpha\tau_{-1}}\frac{\partial}{\partial
\tau_{-1}}$ and
$\tilde{X}_2=A(\tau_{-2})e^{-\alpha\tau_{-2}}\frac{\partial}{\partial
\tau_{-2}}$. They satisfy  relations
$[D_x,\tilde{X}_2]=0$ and $[D_x,\tilde{X}_1]=\mu\tilde{X}_2$. Here
the coefficient $\mu$ is
$$\mu=\alpha A(\tau_{-1})e^{-2\alpha\tau_{-1}+\alpha t}.
$$

\noindent Construct a sequence by taking
$\tilde{X}_1,\tilde{Y}_1,C_2=[\tilde{X}_1,\tilde{Y}_1],C_3=[\tilde{X}_1,C_2],C_k=[\tilde{X}_1,C_{k-1}]$
for $k\geq 3$. One can easily check that
 $$[D_x,C_2]=-\tilde{Y}_1(\mu)\tilde{X}_2+\tilde{X}_1(\lambda_1)\tilde{Y}_0=
b_2\tilde{X}_2+\tilde{X}_1(\lambda_1)\tilde{Y}_0,$$
$$[D_x,C_3]=\tilde{X}_1^2(\lambda_1)\tilde{Y}_0-(C_2+\tilde{X}_1\tilde{Y}_1)(\mu)\tilde{X}_2=
\tilde{X}_1^2(\lambda_1)\tilde{Y}_0+b_3\tilde{X}_2,$$ and for any
$k$ (it can be proved by induction)
$$[D_x,C_k]=\tilde{X}_1^{k-1}(\lambda_1)\tilde{Y}_0+b_k\tilde{X}_2.$$

\noindent Since the characteristic Lie algebra $L_n$ is of finite
dimension then there is a number $N$ such that
\begin{equation}\label{CN+1}C_{N+1}=\mu_NC_N+...+\mu_1\tilde{Y}_1+\mu_0\tilde{X}_1,\end{equation}
where $\tilde{X}_1$, $\tilde{Y}_1$, $C_1$, $C_2$, $\ldots$ are
linearly independent.\\

\noindent Commute both sides of (\ref{CN+1}) with $D_x$ and get
\begin{eqnarray*}
\tilde{X}_1^N(\lambda_1)\tilde{Y}_0+b_{N+1}\tilde{X}_2&=&D_x(\mu_N)C_N+...+D_x(\mu_1)\tilde{Y}_1
+D_x(\mu_0)\tilde{X}_1\\&&+\mu_N\tilde{X}_1^{N-1}(\lambda_1)\tilde{Y}_0+...+\mu_1\lambda_1\tilde{Y}_0+
\{\sum\limits_{k=2}^Nb_k\mu_k\}\tilde{X}_2.
\end{eqnarray*}
\noindent Collect the coefficients before the operators and get
$D_x(\mu_j)=0$ for $j=0,1,...,N$, and
\begin{equation}\label{eqnlambda_1}
(\tilde{X}_1^N-\mu_N\tilde{X}_1^{N-1}-\mu_{N-1}\tilde{X}_1^{N-2}-...-\mu_1)\lambda_1=0.
\end{equation}
\noindent Introduce new variables $\eta, \eta_{-1}$ as solutions
of the following ordinary differential equations
\begin{equation}\label{etaandeta_(-1)}
\frac{d\tau_{-1}}{d\eta}=A(\tau_{-1})e^{-\alpha\tau_{-1}},\quad
\frac{d\tau_{-2}}{d\eta_{-1}}=A(\tau_{-2})e^{-\alpha\tau_{-2}}.
\end{equation}
\noindent Thus our vector fields are rewritten as
$$\tilde{X}_1=\frac{\partial}{\partial \eta},\quad \tilde{X}_2=\frac{\partial}{\partial \eta_{-1}},
\quad \tilde{Y}_0=A(\tau)\frac{\partial}{\partial t_1},$$
$$\tilde{Y}_1=e^{\alpha\tau_{-1}}\frac{\partial}{\partial\eta}+A(\tau_{-1})\frac{\partial}{\partial \tau_{+}}
+D_x(A(\tau_{-1}))\frac{\partial}{\partial t_x}+...\,.$$

\noindent By looking at the projection on
$\frac{\partial}{\partial \eta}$ we get an algebra generated by
$\frac{\partial}{\partial \eta}$ and $e^{\alpha
\tau_{-1}}\frac{\partial}{\partial\eta}$ containing all possible
commutators and all possible linear combinations with constant
coefficients. Due to Lemma  \ref{Lemma6}, we
get that $e^{\alpha\tau_{-1}}$ can be only one of the forms
\begin{enumerate}
\item[a)]$e^{\alpha\tau_{-1}}=c_1e^{\beta\eta}+c_2e^{-\beta\eta}+c_3$,
\item[b)]$e^{\alpha\tau_{-1}}=c_1\eta^2+c_2\eta+c_3$,
\end{enumerate}
where $\beta$, $c_1$, $c_2$, $c_3$ are some constants.

 The
equation
$A(\tau_{-1})=\frac{1}{\alpha}\frac{d}{d\eta}e^{\alpha\tau_{-1}}$
implies that \\
 in case a) we have $A(\tau_{-1})=(\beta/\alpha)(c_1e^{\beta\eta}-c_2e^{-\beta\eta})$,  or the same
\begin{equation}\label{function3a}
A^2(\tau)= \frac{\beta^2}{\alpha^2}\{(e^{\alpha \tau }-c_3)^2-4c_1c_2\}\, ,
\end{equation}
 and\\
 in case b) we have $A(\tau_{-1})=(1/\alpha)(2c_1\eta+c_2)$, or the same,
 \begin{equation}\label{function3b}
 A^2(\tau)=\frac{4c_1}{\alpha^2}e^{\alpha \tau}+\frac{c_2^2-4c_1c_3}{\alpha^2}.
 \end{equation}
In addition to the operators $\tilde{X}_1, \tilde{X}_2, \tilde{Y}_0,
\tilde{Y}_1$ introduced above we will use
$\tilde{Y}_2=A(\tau_{-2})D^{-1}(Y_1f)\partial_{t_x}+A(\tau_{-2})D^{-1}(Y_1f_x)\partial_{t_{xx}}+...$
defined as $\tilde{Y}_2=A(\tau_{-2})Y_2$. It satisfies the
commutativity relation
\begin{equation}\label{commutativitytildeY_2}
[D_x,\tilde{Y}_2]=\lambda\tilde{Y}_1+\xi\tilde{Y}_0+\nu\tilde{X}_1,
\end{equation}
where
\begin{equation}\label{xi}
\xi=-\frac{A(\tau_{-2})}{A(\tau)}D^{-1}(Y_1f)=
-\frac{A(\tau_{-2})}{A(\tau)}\{(-A'(\tau_{-1})+\alpha
A(\tau_{-1}))e^{-\alpha\tau_{-1}}+A'(\tau_{-2})e^{-\alpha\tau_{-2}-\alpha\tau_{-1}}
\}e^{\alpha t}.
\end{equation}
$$\lambda=-\frac{A(\tau_{-2})}{A(\tau_{-1})}D^{-1}(Y_1f) \qquad {\mbox{and }} \qquad
\nu=-\lambda e^{\alpha\tau_{-1}}\, . $$

\begin{lem}\label{c1c2=0}\quad (1) Equation
$t_{1x}=t_x+\frac{\beta}{\alpha}(e^{\alpha \tau }-c_3)e^{\alpha
t}$ admits a nontrivial $n$-integral if and only if $c_3=\pm 1$.\\
(2)  Equation $t_{1x}=t_x+c_5e^{\alpha t}$, $c_5\ne 0$ does not
admit a nontrivial $n$-integral.
\end{lem}

\noindent{\textbf{ Proof.}} In this case the equation
$\tau_x=A(\tau)e^{\alpha t}$ is reduced by evident scaling of $x$
and $t$ to
$$t_{1x}=t_x +e^t, \qquad {\mbox{or}}\qquad t_{1x}=t_x+e^{t_1}+\varepsilon e^t.$$
By induction on $n$ one can easily see that for the equation
$t_{1x}=t_x +e^t$, the basic vector fields $Y_n$ are
$$ Y_1=\frac{\partial}{\partial t}, $$
$$
Y_n=e^{t_{-(n-1)}}\frac{\partial}{\partial t_x}
+e^{t_{-(n-1)}}(t_x-e^{t_{-(n-1)}})\frac{\partial}{\partial
t_{xx}} +\ldots
$$
Since these vector fields $Y_n$, $n\geq 1$,  are linearly
independent then equation $t_{1x}=t_x+e^t$ does not admit a
nontrivial $n$-integral.

\noindent For equation $t_{1x}=t_x+e^{t_1}+\varepsilon e^t$, the
basic vector fields $Y_n$ are
$$
Y_1=\frac{\partial }{\partial t}+e^t\frac{\partial}{\partial
t_x}+e^t(t_x+e^t)\frac{\partial}{\partial t_{xx}}+\ldots,
$$
$$
Y_n=(\varepsilon +1)e^{t_{-(n-1)}}\frac{\partial}{\partial
t_x}+(\varepsilon +1)e^{t_{-(n-1)}}(t_x+(1-\varepsilon
)e^{t_{-(n-1)}})\frac{\partial}{\partial t_{xx}}+\ldots
$$
One can see that vector fields $Y_n$, $n\geq 1$,  are linearly
independent if $\varepsilon\ne \pm 1$. Therefore, if
$\varepsilon\ne \pm 1$, equation $t_{1x}=t_x+e^{t_1}+\varepsilon
e^t$ does not admit a nontrivial $n$-integral. If $\varepsilon
=-1$, the equation becomes $t_{1x}=t_x+e^{t_1}-e^t$, and one of
its $n$-integrals is $I=t_x-e^t$. If $\varepsilon =1$, the
equation becomes $t_{1x}=t_x+e^{t_1}+e^t$, and one of its
$n$-integrals is $I=2t_{xx}-t_x^2-e^{2t}$. $\Box$

\begin{lem}\label{Y2notexpressed}Let equation
$t_{1x}=t_x+A(t_1-t)e^{\alpha t}$ with \\
(a) $A^2(\tau)= \frac{\beta^2}{\alpha^2}\{(e^{\alpha \tau }-c_3)^2-4c_1c_2\}$,  or\\
(b) $A^2(\tau)=\frac{4c_1}{\alpha^2}e^{\alpha \tau}+\frac{c_2^2-4c_1c_3}{\alpha^2}$, \\
admit a nontrivial $n$-integral.
Then \\
in case (a), we have,
 $A(t_1-t)=\frac{\beta}{\alpha}\sqrt{(e^{\alpha(t_1-t)}-c_3)^2-c_3^2+1}$, where $c_3$ is an arbitrary constant, \\
and\\
in case (b), we have,
 $A(t_1-t)=ce^{\frac{\alpha}{2}(t_1-t)}$, where $c$ is an arbitrary constant. \\
In cases (a) and  (b) the corresponding $n$-integrals are
$I=\frac{\alpha}{2}t_x^2-t_{xx}+ \frac{\alpha}{2}e^{2\alpha t}$ and
$I=-\frac{\alpha}{2}t_x^2+t_{xx}$.

 \end{lem}

\noindent{\textbf{ Proof.}} Note that
$$D_x \rho =\lambda, \qquad {\mbox{where}} \qquad \rho =-\frac{A(\tau_{-2})}{A(\tau_{-1})}-e^{\alpha
\tau_{-2}}.
$$
This implies that the vector field
$$
R_2=\tilde{Y}_2-\rho \tilde{Y}_1,
$$
satisfies very simple and convenient relation
$$
[D_x, R_2]=\tilde{\xi}\tilde{Y}_0+\nu \tilde{X}_1\, \qquad
\tilde{\xi}=-\frac{A(\tau_{-2})}{A(\tau)}D^{-1}(Y_1f)-\rho
\lambda_1, \qquad \nu=e^{\alpha
\tau_{-1}}\frac{A(\tau_{-2})}{A(\tau_{-1})}D^{-1}(Y_1f).
$$
Study now the sequence
$$
R_{j+1}=[\hat{X}, R_j], \quad j\geq 2, \qquad {\mbox{where }} \qquad \hat{X}=\tilde{X}_1+e^{-\alpha \tau_{-1}}
\tilde{X_2}\, .
$$
Direct calculations show that
\begin{equation}\label{Rn}
[D_x,R_n]=\hat{X}^{(n-2)}(\tilde{\xi})\tilde{Y}_0+\hat{X}^{(n-2)}(\tilde{\nu})\tilde{X}_1 +b_n\tilde{X}_2\, .
\end{equation}
Since $\tilde{X_1}$, $\tilde{X_2}$, $\tilde{Y_0}$, $R_2$ are
linearly independent, then there exists a  number $N\geq 2$ such
that
$$
R_{N+1}=\mu_N R_N+\mu_{N-1}R_{N-1}+\ldots
\mu_2R_2+\mu_1\tilde{X}_1
$$
and
\begin{equation}\label{compare}
[D_x,R_{N+1}]=[D_x,\mu_N R_N+\mu_{N-1}R_{N-1}+\ldots
\mu_2R_2+\mu_1\tilde{X}_1]\, .
\end{equation}
We use $[D_x, \tilde{X}_1]=\alpha A(\tau_{-1})e^{-2\alpha
\tau_{-1}+\alpha t}\tilde{X}_2$, $[D_x, \tilde{X_2}]=0$ and
(\ref{Rn}) to compare the coefficients before linearly independent
vector fields $R_k$ and $\tilde{Y_0}$ in (\ref{compare}). We have,
$D_x(\mu_k)=0$, $k=2,3,\ldots, N$,  and
\begin{equation}\label{tildexi}
\hat{X}^{(N-1)}(\tilde{\xi})=\mu_N\hat{X}^{(N-2)}(\tilde{\xi})+\ldots
+ \mu_2\tilde{\xi}\, .
\end{equation}
Under the change of variables
$$
\eta=z, \qquad \eta_{-1}=z_{-1}-q(z), \qquad \frac{\partial
q(z)}{\partial z}=-e^{-\alpha \tau_{-1}}\, ,
$$
equation (\ref{tildexi}) is reduced to
\begin{equation}\label{final}(
D_z^{N-1}-\mu_ND_z^{N-2}-\ldots -\mu_2)\tilde{\xi}=0,
\end{equation}
where $\mu_k=\mu_k(\tau_{-1}, \tau_{-2})=\mu_k(z, z_{-1})$.
Since $D_x(z_{-1})=0$,  $D_x(z)=e^{\alpha t}\ne 0$ and $0=D_x(\mu_k)=D_{z_{-1}}(\mu_k)D_x(z_{-1})+D_z(\mu_k)D_x(z)$, then coefficients $\mu_k$ do not depend on variable $z$ .
Since, due to (\ref{final}),
$$
\tilde{\xi}=-\frac{A(\tau_{-2})}{A(\tau)}e^{-\alpha\tau_{-1}}e^{\alpha t}
\{-A'(\tau_{-1})+\alpha A(\tau_{-1})+A'(\tau_{-2})e^{-\alpha\tau_{-2}}\}
$$
$$+
\frac{A(\tau_{-2})}{A(\tau)}e^{\alpha t}\{A'(\tau)-\alpha A(\tau)-A'(\tau_{-1})e^{-\alpha \tau_{-1}}\}
$$
$$+\frac{A(\tau_{-1})}{A(\tau)}e^{\alpha \tau_{-2}}e^{\alpha t}\{A'(\tau)-\alpha A(\tau)-A'(\tau_{-1})e^{-\alpha \tau_{-1}}\}\,
$$
is a quasi-polynomial in $z=\eta$ for any $\tau$ and $t$,  then
$\frac{d}{d\tau}(\tilde{\xi}A(\tau)e^{-\alpha t})$ is a
quasi-polynomial as well. Hence we have,
$$
(A''(\tau)-\alpha A'(\tau))\{A(\tau_{-2})+A(\tau_{-1})e^{\alpha
\tau_{-2}}\}
$$
is a quasi-polynomial in $z$, which is possible only if\\
$A''(\tau)-\alpha A'(\tau)=0$, or
$A(\tau_{-2})+A(\tau_{-1})e^{\alpha \tau_{-2}}$ is a
quasi-polynomial in $z$.\\
 In case  (a) we have,
$$
A''(\tau)-\alpha A'(\tau)=-\alpha \beta c_4 \frac{e^{2\alpha \tau}}{(\sqrt{(e^{\alpha\tau}-c_3)^2-c_4})^3}\, ,\qquad c_4=4c_1c_2,
$$
and in case (b) we have
$$
A''(\tau)-\alpha A'(\tau)=-4c_1^2\alpha^{-2}e^{2\alpha \tau }\left(\frac{4c_1}{\alpha^2}e^{\alpha \tau }+\frac{c_2^2-4c_1c_3}{\alpha^2}\right)^{-3/2}
$$
Therefore, $A''(\tau)-\alpha A'(\tau)=0$ if $c_1c_2=0$ in case (a)
and if $c_1=0$ in case (b).  Both these cases are considered in Lemma \ref{c1c2=0}.

It follows from $\frac{dq}{dz}=-e^{-\alpha \tau_{-1}}$ that, in case (a), if $r=
\sqrt{c_3^2-4c_1c_2}\ne 0$, then
$$
q(\eta)=-\frac{1}{\beta r}\ln\left|\frac{e^{\beta \eta}-p_1}{e^{\beta \eta}-p_2}\right|, \qquad p_1=\frac{-c_3+r}{2c_1}, \quad p_2=\frac{-c_3-r}{2c_1},
$$
and if
$r=
\sqrt{c_3^2-4c_1c_2}= 0$, then
$$
q(\eta)=\frac{1}{c_1\beta (e^{\beta \eta}-p_1)}.
$$
In case (b), if $r_1=\sqrt{c_2^2-4c_1c_3}\ne 0$, then
$$
q(\eta)=-\frac{1}{\beta r_1}\ln\left|\frac{\eta-p_1^*}{\eta-p_2^*}\right|, \qquad p_1^*=\frac{-c_2+r_1}{2c_1}, \quad p_2^*=\frac{-c_2-r_1}{2c_1},
$$
and if $r_1=\sqrt{c_2^2-4c_1c_3}= 0$, then
$$
q(\eta)=\frac{1}{c_1\beta(\eta-p_1^*)}\, .
$$

In case (a) we have,
$$
\frac{\alpha}{\beta}(A(\tau_{-2})+A(\tau_{-1})e^{\alpha
\tau_{-2}})=c_1e^{\beta\eta_{-1}}-c_2e^{-\beta\eta_{-1}}+(c_1e^{\beta\eta}-c_2e^{-\beta\eta})(c_1e^{\beta\eta_{-1}}+c_2e^{-\beta\eta_{-1}}+c_3)
$$
$$
=c_1e^{\beta \eta_{-1}}(c_1e^{\beta \eta}-c_2 e^{-\beta \eta}+1)+c_2e^{-\beta \eta_{-1}}
(c_1e^{\beta \eta}-c_2e^{-\beta \eta}-1) +c_3c_1e^{\beta \eta}-c_3c_2e^{-\beta \eta}
$$
$$
=c_1e^{\beta z_{-1}-\beta q(z)}(c_1e^{\beta z}-c_2 e^{-\beta
z}+1)+c_2e^{-\beta z_{-1}+\beta q(z)} (c_1e^{\beta
z}-c_2e^{-\beta z}-1) +c_3c_1e^{\beta z}-c_3c_2e^{-\beta
z}.
$$
One can see that $A(\tau_{-2})+A(\tau_{-1})e^{\alpha
\tau_{-2}}$ is a quasi-polynomial in case (a) only if  $r=\sqrt{c_3^2-4c_{1}c_{2}}=\pm 1$.
If $r=\pm 1$, function $A(t_1-t)$ becomes
$\frac{\beta}{\alpha}\sqrt{(e^{\alpha(t_1-t)}-c_3)^2-c_3^2+1}$, where $c_3$ is an arbitrary constant, \\
and one of
 $n$-integrals for $t_{1x}=t_x+\frac{\beta}{\alpha}e^{\alpha t}\sqrt{(e^{\alpha(t_1-t)}-c_3)^2-c_3^2+1}$ is
$I=\frac{\alpha}{2}t_x^2-t_{xx}+ \frac{\alpha}{2}e^{2\alpha t}$.

In case (b) direct calculations show that,
$$
A(\tau_{-2})+A(\tau_{-1})e^{\alpha \tau_{-2}}=Q(z)+P(z, z_{-1})+J(z,z_{-1}),
$$
where $Q(z)$ is some function depending only on $z$, $P(z, z_{-1})$ is a polynomial function of two variables,  and
$$
J(z,z_{-1})=-\frac{2c_1}{\alpha}z_{-1}q(z)(2c_1 z+c_2).
$$
Since $A(\tau_{-2})+A(\tau_{-1})e^{\alpha \tau_{-2}}-P(z, z_{-1})=Q(z)+J(z,z_{-1})$
is a quasi-polynomial in $z$, then
$$\frac{\partial (Q(z)+J(z,z_{-1}))}{\partial z_{-1}}=\frac{2c_1}{\alpha}q(z) (2c_1 z+c_2)
$$
is also a quasi-polynomial in $z$, which is possible only if $r_1=\sqrt{c_2^2-4c_1c_3}=0$.
If $r_1=0$ we have
$A(t_1-t)=ce^{\frac{\alpha}{2}(t_1-t)}$, where $c$ is an arbitrary constant, and  the corresponding $n$-integral is
$I=-\frac{\alpha}{2}t_x^2+t_{xx}$. $\Box$

\subsection{Case 4) $t_{1x}=t_x+c_4(e^{\alpha t_1}-e^{\alpha
t})+c_5(e^{-\alpha t_1}-e^{-\alpha t})$}

It is clear that this equation has a nontrivial $n$-integral which
is $I=t_x-c_4e^{\alpha t}+c_5e^{-\alpha t}$. It satisfies the
equation $DI=I$ since $DI=t_{1x}-c_4e^{\alpha t_1}+c_5e^{-\alpha
t_1}=I$.

\section{Conclusion}
In the article we studied differential-difference equations of the
form (\ref{dhyp}) from the Darboux integrability point of view. We
showed that all Darboux integrable chains are connected with one
another by Cole-Hopf type differential substitutions. The problem of
classification of Darboux integrable chains is studied by reducing
it to an adequate algebraic form. We use the fact  that the chain
(\ref{dhyp}) is Darboux integrable if and only if its characteristic
Lie algebras $L_x$ and $L_n$ both are of finite dimension to obtain the  complete
list of Darboux integrable chains of the particular form
$t_{1x}=t_x+d(t,t_1)$.

\section*{Acknowledgments}
This work is partially supported by the Scientific and
Technological Research Council of Turkey (T\"{U}B{\.{I}}TAK). One
of the authors (IH) thanks Russian Foundation for Basic
Research (RFBR) (grants $\#$ 09-01-92431KE-a, $\#$
08-01-00440-a, and $\#$ 07-01-00081-a).

\end{document}